# Superconductivity in Metal-mixed Ion-Implanted Polymer Films

A.P. Micolich [a]
*School of Physics, University of New South Wales, Sydney NSW 2052, Australia.*

E. Tavenner and B.J. Powell
*Physics Department, University of Queensland, Brisbane QLD 4072, Australia*

A.R. Hamilton
*School of Physics, University of New South Wales, Sydney NSW 2052, Australia.*

M.T. Curry and R.E. Giedd
*Center for Applied Science and Engineering, Missouri State University, Springfield MO 65804.*

P. Meredith [b]
*Physics Department, University of Queensland, Brisbane QLD 4072, Australia*

Ion-implantation of normally insulating polymers offers an alternative to depositing conjugated organics onto plastic films to make electronic circuits. We used a 50 keV nitrogen ion beam to mix a thin 10 nm Sn/Sb alloy film into the sub-surface of polyetheretherketone (PEEK) and report the low temperature properties of this material. We observed metallic behavior, and the onset of superconductivity below 3 K. There are strong indications that the superconductivity does not result from a residual thin-film of alloy, but instead from a network of alloy grains coupled via a weakly conducting, ion-beam carbonized polymer matrix.

Since the discovery of electrical conductivity in conjugated polymers,[1] intense international effort has focused on the development of electronic devices and integrated circuits on plastic films.[2,3] Such 'plastic electronics' promise significant advantages over existing technologies (e.g., Si microelectronics), including simpler and more efficient processing, reduced production and materials costs, mechanical flexibility/robustness and the possibility of large area (~m$^2$) production.[3] Conducting polymers such as heavily doped polyacetylene can exhibit a metallic state.[4,5] However, metallic behavior in organic polymers is not common, and the underlying physics is not well understood.[6] Ion implantation of normally insulating polymers offers an alternative to depositing metals or conjugated organics onto plastic films to make electronic circuits. A key advantage of this approach is that ion implantation techniques are widely used in the semiconductor industry and could be adapted for processing electronic circuits based on plastic substrates.

Numerous studies of ion-implantation into insulating polymer substrates have reported increased electrical conductivity due to carbonization of the polymer by the ion beam.[7-10] However, despite showing conductivities up to ~300 S/cm, these materials are insulators, exhibiting increasing resistivity with decreasing temperature. Achieving metallic conductivity in ion-implanted polymers is a long-standing problem. Recently, implantation of polyetheretherketone (PEEK – chemical structure shown inset to Fig. 1(c)) using a metallic Sn ion-beam was explored.[11] However, this resulted in a maximally implanted ion content insufficient for metallic conductivity due to a self-limiting sputtering process.[12] One way to overcome this problem is to deposit a thin metal layer on the polymer substrate and then use an ion-beam to 'mix' this metal into the polymer sub-surface – a process known as 'metal-mixing'.[13] The rationale behind metal-mixing is that lower mass inert ions can be used, reducing the sputtering, while the metal layer ensures that after implantation, sufficient metal atoms have been mixed into the polymer to allow metallic conductivity.

In this paper, we report the results of a low-temperature electrical study of this Sn:Sb/PEEK metal mixed system. In addition to a weak metallic temperature dependence, we observe the onset of superconductivity at temperatures $T$ < 3 K. Previous studies of organic superconductor-polymer blends (e.g., $\beta$-(ET)$_2$I$_3$ microcrystals in polycarbonate) have observed a partial Meissner effect.[14] Here we report a plastic material that shows a complete transition to a robust zero electrical resistance state, suggesting potential technological applications for the ion-beam processed polymer that are not possible with superconducting-polymer blends.

Our samples were prepared by evaporating a ~10 nm layer of 95% Sn : 5% Sb alloy onto a 0.1 mm thick PEEK film, and subsequently implanting the metallized surface to doses of 10$^{16}$ (sample A) and 10$^{15}$ (sample B) ions/cm$^2$ with a 50 keV N$^+$ beam. We also prepared a control sample consisting of bare PEEK implanted with 50 keV N$^+$ ions to $10^{16}$ ions/cm$^2$ (sample C). The 5% Sb plays an essential role in the conductivity of samples A and B. Samples prepared with 100% Sn are strongly insulating, even with Sn film thicknesses as high as 40 nm. It is well known that the addition of Sb to Sn inhibits its transformation from metallic white allotrope to the insulating grey allotrope.[15,16] However, it is not yet clear if/how the 5% Sb fraction affects the post-implant structure of the samples – we note that an underlayer of Sb is often used in quench condensed film studies to control the length scale of disorder (between atomic and mesoscopic) in the system.[19]

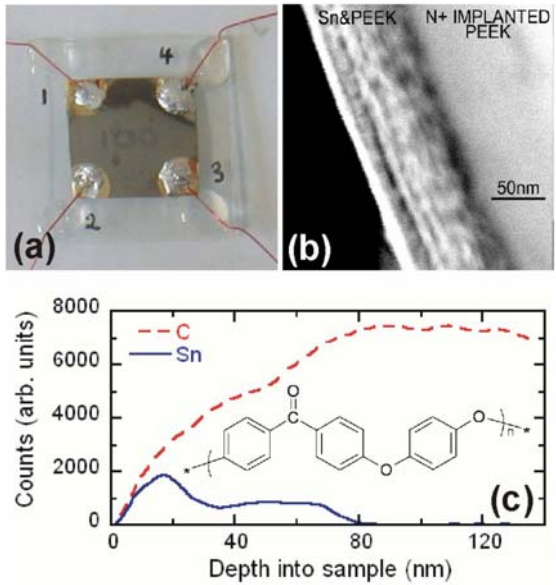

Fig. 1: (a) A photograph of a metal mixed sample which consists of a 10 nm layer of Sn:Sb alloy on a 0.1mm thick PEEK film that was subsequently implanted with a 50keV N$^+$ beam to a dose of $10^{16}$ ions/cm$^2$. The sample is 15mm square with 50nm Ti / 50nm Au contacts evaporated in the corners and wires attached using InAg solder. (b) Cross-sectional STEM image of a metal mixed sample, and (c) corresponding energy dispersive x-ray analysis (EDAX) profile showing the relative concentrations of Sn (solid line) and C (dashed line) in the near surface of the PEEK. (b) and (c) demonstrate that implantation has mixed the initial 10 nm Sn:Sb alloy layer over ~75 nm of the sub-surface region of the PEEK film. The chemical structure of PEEK is inset to (c).

To demonstrate that the ion-beam thoroughly mixes the Sn/Sb alloy film into the PEEK surface, in Fig. 1(b) we show a cross-sectional scanning transmission electron microscopy (STEM) image of the near-surface region of a sample nominally identical to sample A. Clear structural differences are evident between the implant-mixed region, extending ~75 nm into the sample, and the bulk polymer. The accompanying EDAX elemental analysis profile, shown in Fig. 1(c), confirms that the original 10 nm of alloy is now distributed over more than seven times its original volume. Previous X-ray Photoelectron Spectroscopy (XPS) studies of this material[11] show that implantation induces three key chemical changes relative to an unimplanted sample: a) the number of Sn-Sn bonds is reduced by a factor of 4 while the Sn-C bond content is increased from < 0.1 % to ~5 %; b) there is a net decrease in Sn-Sn bond content in the first ~8.5 nm (the region probed by XPS) of the sample, due to a combination of mixing the Sn deeper into the sample and sputtering by the incident energetic N$^+$ ions, and; (c) the graphitic carbon content is increased from < 0.1 % to ~27 %, consistent with previous studies.[7-10] The XPS data also indicates that much of the 5% Sb fraction is oxidized during implantation, with net Sb loss due to sputtering.[11] Combined, these chemical and structural findings strongly suggest that the alloy film has been thoroughly mixed into the PEEK.

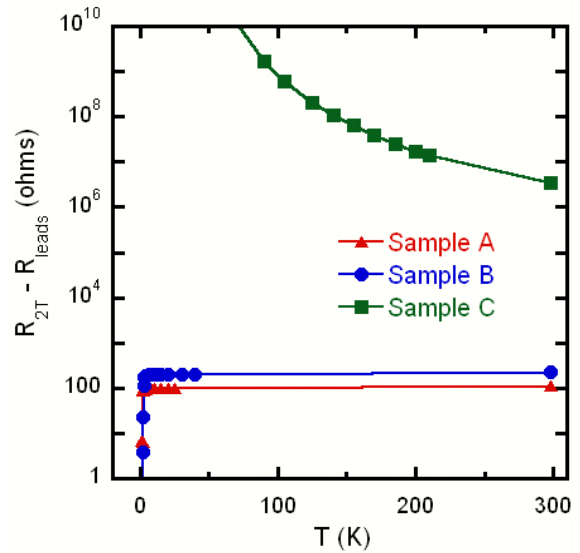

Fig. 2: Resistance (two-terminal minus temperature-dependent lead resistance) $R_{2T}$-$R_{leads}$ on a log scale vs. temperature $T$ for sample A (10 nm Sn:Sb on PEEK, $10^{16}$ ions/cm$^2$ 50 keV N$^+$), sample B (10 nm Sn:Sb on PEEK, $10^{15}$ ions/cm$^2$ 50 keV N$^+$) and sample C (bare PEEK, $10^{16}$ ions/cm$^2$ 50 keV N$^+$). The addition of the thin Sn:Sb layer prior to implantation leads to a metallic temperature dependence in contrast to previous samples,[7-10] such as sample C, which show an insulating temperature dependence. A sharp drop in resistance is apparent in samples A and B at low $T$.

After implantation, the samples were prepared for electrical characterization. Four 5mm diameter contacts were deposited at the corners of the 15mm square samples, to form a quasi-Van der Pauw measurement configuration (see Fig. 1(a)). Contacts consisting of 50 nm Ti / 50 nm Au were deposited by thermal evaporation through a shadow mask. After contact deposition, the samples were mounted on 25mm square glass slides and insulated Cu wires were attached using low melting point InAg solder.

Electrical measurements were performed in an Oxford Instruments VTI system, which allowed temperatures $T$ between 1.2 K and 200 K and magnetic fields $B$ up to 10 T. The d.c. electrical resistance $R$ of the samples was measured using a Keithley 2400 Source-Measure Unit in both two- and four-terminal modes. We commenced by measuring the two-terminal electrical resistance $R_{2T}$ vs $T$ for the three samples. In order to establish the true temperature-dependence of the sample, we obtained the temperature-dependent lead resistance $R_{leads}$ by simultaneously measuring two nominally identical leads shorted together at the bottom of the cryostat, and subtracted this from $R_{2T}$ to give the data shown in Fig. 2.

Two features are evident from this data (Fig. 2). Firstly, we observe a weak metallic temperature dependence (i.e., $R$ decreasing with decreasing $T$) in the metal-mixed samples A and B. This is in contrast to sample C (the $N^+$ implanted sample), where we observe very strong insulating behavior consistent with previous studies.[7-10] Secondly, the data in Fig. 2 suggest that samples A and B, although metallic, are highly disordered. The level of disorder is typically characterized by the residual resistance ratio (RRR) defined as $\rho(300K)/\rho(T_c^+)$, where $T_c^+$ is a temperature slightly above $T_c$, and for samples A and B we find a maximum RRR of 1.2,

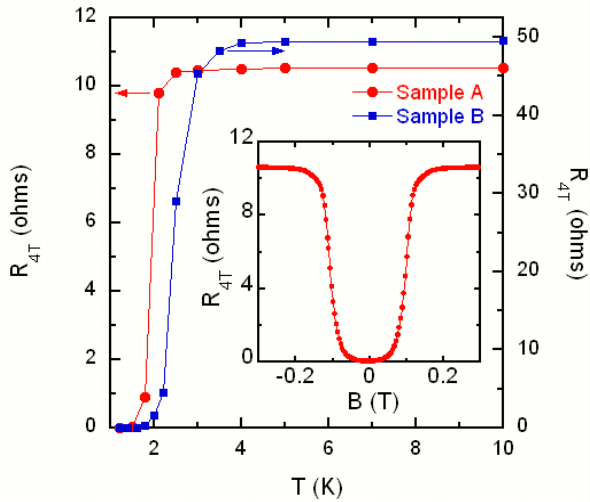

Fig. 3: Four-terminal resistance $R_{4T}$ vs $T$ for sample A (left axis) and B (right axis) demonstrating the presence of a zero resistance (i.e., superconducting) state at low $T$. The critical temperatures $T_c$ are 1.9K and 2.4K for samples A and B, respectively. (inset) $R_{4T}$ vs magnetic field $B$, oriented perpendicular to the plane of the plastic film, at $T = 1.2$ K for sample A, demonstrating a field-induced transition to a normal state. The measured critical field $B_c$ is 0.12 T.

indicative of a highly disordered material.

The data in Fig. 2 show a very sharp drop in the resistance at $T < 3K$, typical of that observed in superconductors. We used four-terminal measurements, shown in Fig. 3, to further explore the possibility of superconductivity. We again observe a clear metallic temperature dependence that culminates in a sharp drop, this time to a zero resistance state ($R_{4T} < 0.01$ $\Omega$, our instrument sensitivity limit) at a critical temperature $T_c$ of 1.9 K and 2.4 K for samples A and B, respectively. The observed electrical behavior is repeatable after thermal cycling to room temperature, and reproducible (quantitatively similar) in nominally identical samples. We also find that the metal-mixed layer does not delaminate even after several cryogenic cycles and the implanted material shows significant durability. We have repeated our measurements over a seven month period with little change or degradation of the electronic properties beyond a slight (< 10%) increase in the normal resistance over this period, despite simply storing these samples in a plastic box under ambient conditions. We have also obtained magnetic field dependence measurements on sample A (field oriented perpendicular to the 2D plane of the sample) to establish the critical magnetic field $B_c$. As shown in Fig. 3 (inset), we see a field-induced superconducting-normal transition at $B_c = 0.12$ T, and as expected, $B_c$ decreases with increasing $T$. The typical critical current $I_c$ for this material is of order 1 mA.

At first sight, the observation of superconductivity in these samples may not be surprising – Sn is an elemental superconductor with $T_c = 3.7$ K and $B_c = 30$ mT, and Sb becomes superconducting under pressure.[16] We find that a 20 nm thick film of Sn:Sb on PEEK (unimplanted) shows a very sharp transition with $T_c = 3.7$ K and $B_c \sim 300$ mT (i.e., neither the Sb impurities nor geometrical effects reduce $T_c$). While the increased $B_c$ is expected due to the thin film geometry, the suppressed $T_c$ observed in samples A and B is surprising[17] if the observed superconductivity results from bulk alloy. Hence we instead consider two likely models for the superconductivity: a) a residual layer of alloy thin enough to not only increase $B_c$ relative to bulk Sn, but also suppress $T_c$, and; b) a granular alloy system where the weakly conducting ion-beam modified PEEK matrix acts as a weak link between Josephson- or proximity-effect coupled grains.[18] We also note a more exotic possibility (which will not be considered further here), which is a metal-carbon 'eutectic' where the Sn and C are intimately mixed at the molecular level and the Sn acts mainly by doping the hydrocarbon.

We will now critically assess the two models presented above against the experimental data. Firstly, the correct model must explain why $T_c$ is significantly suppressed. The suppression of $T_c$ has previously been observed in quench-condensed films grown at low temperatures[19,20] ($\leq 4K$), where higher disorder (indicated by a higher normal resistance $R(T_c^+)$) leads to reduced $T_c$.[20] Considering the data in Fig. 3 from this viewpoint, we would expect that a lower implant dose would lead to less disorder and consequently a lower $R(T_c^+)$. However, as Fig. 3 shows, the lower dose sample B actually has a $R(T_c^+) \sim 4$ times higher than sample A. The relationships between the $R(T_c^+)$ and $T_c$ values measured from the data presented in Fig. 3 clearly contradict the behavior observed in quench-condensed films[20] – *sample A, which has the lower $R(T_c^+)$, and thus the least disorder, actually has the lower $T_c$.* Furthermore, $T_c$ suppression in quench condensed films only occurs when the resistance is of order the quantum of resistance for electron pairs (i.e., $R(T_c^+) \sim h/4e^2 \sim 6450$ $\Omega/\square$), which is more than two orders of magnitude larger than $R(T_c^+)$ in either sample A or B (see Fig. 3), which were measured in the standard square geometry (see Fig. 1(a)).

We now consider a network of Sn:Sb alloy granules coupled via a weakly conducting, carbonized polymeric matrix. Firstly, this possibility is consistent with our structural analysis, in particular, the decrease in Sn-Sn bonds and the increase in Sn-C bonds indicated by XPS. The granular model also provides an explanation for why both the $R(T_c^+)$ and $T_c$ of sample A, which has the higher implant dose, are lower than those of sample B. Considering $R(T_c^+)$ first, in the normal phase, $R$ is dominated by inter-grain hopping and increases with the grain separation. Sample A, which has the higher implant dose, would be expected to contain smaller alloy grains with a smaller inter-grain separation (i.e., better mixing), and hence should have a lower $R(T_c^+)$, as observed in Fig. 3. Considering $T_c$, supposing that the grains are small enough that they do undergo $T_c$ suppression, then smaller grains would be expected to have a lower $T_c$, and hence the higher dose sample A would have the lower $T_c$, as also observed in the data in Fig. 3. While we currently believe that the granular hypothesis gives the most natural explanation for our data, experiments that establish the micro/nano-scale structure of the ion-beam mixed region are required to confirm this hypothesis. We also plan to explore metal-mixing of other elements (e.g., Nb) with the goal of increasing $T_c$.

In conclusion, we have used ion-implantation to mix a thin Sn:Sb alloy layer into a PEEK surface to significantly enhance the conductivity. Using this approach we observe not only metallic behavior, but also the onset of superconductivity at $T < 3$ K. Our combined structural and electrical data strongly suggest that the metallic conductivity and superconductivity arise from a network of alloy grains coupled through a weakly conducting, ion-beam carbonized, polymer matrix, as opposed to a residual thin film of alloy.

We thank A. Ardavan, A. Briggs, J. Brooks and particularly R. McKenzie for helpful conversations, staff at the Centre for Microscopy and Microanalysis (UQ) for microscopy/surface analysis. We acknowledge financial support from ARC, AINSE and ONR. APM acknowledges an ARC Postdoctoral Fellowship.

[1] C. K. Chaing et al., *Phys. Rev. Lett.* **39**, 1098 (1977).
[2] G. H. Gelinck et al., *Nature Materials* **3**, 106 (2004).
[3] D. Voss, *Nature* **407**, 442 (2000).
[4] T. Ishiguro et al., *Phys. Rev. Lett.* **69**, 660 (1992).
[5] R. S. Kohlman et al., *Phys. Rev. Lett.* **78**, 3915 (1997).
[6] R. S. Kohlman and A. J. Epstein, in *Handbook of Conducting Polymers, 2nd Ed.*, edited by T. Skotheim, R. Elsenbaumer, J. Reynolds (Marcel Dekker, New York, 1998).
[7] S.R. Forrest et al., *Appl. Phys. Lett.* **41**, 708 (1982).
[8] T. Hioki et al., *Appl. Phys. Lett.* **43**, 30 (1983).
[9] J.A. Osaheni et al., *Macromol.* **25**, 5828 (1992).
[10] V.C. Long, S. Washburn, X.L. Chen, S.A. Jenekhe, *J. Appl. Phys.* **80**, 4202 (1996).
[11] E. Tavenner et al., *Synth. Met.* **145**, 183 (2004).
[12] The momentum of the incident ion-beam results in sputtering of deposited ions from the surface, the rate increasing with incident ion mass. As ions accumulate in the polymer, the probability of one being sputtered back out of the polymer also increases. Combined, these two processes lead to a maximal implanted ion density once the sputtering rate equals the implantation current.
[13] Y.Q. Wang et al., *Nucl. Inst. Meth. Phys. Res. B* **127/128**, 710 (1997).
[14] A. Tracz et al., *Synth. Met.* **120**, 849 (2001).
[15] M.M. El-Bahay, M.E. El-Mossalamy, M. Madhy and A.A. Bahgat, *Phys. Stat. Sol. (a)* **198**, 76 (2003).
[16] N.W. Ashcroft and N.D. Mermin, *Solid State Physics* (Thomson, Singapore, 1976).
[17] K. Maki, *Superconductivity* (Ed. R.D. Parks, Marcel Dekker, New York, 1969).
[18] For example, G. Deutsher, *The Physics of Superconductors*, (Ed. K.H. Bennemann and J.B. Ketterson, Springer, Berlin, 2004).
[19] A.M. Goldman and N. Markovic, *Phys. Today* **51**, 39 (1998).
[20] J.M. Valles Jr., R.C. Dynes, and J.P. Garno, *Phys. Rev. B* **40**, 6680 (1989).

**EPAPS Additional Material**

**Detailed Experimental Methods and Sample Preparation**

**Fabrication:** Thin films (0.1 mm) of polyetheretherketone (PEEK), obtained from the Goodfellow Company, were washed with methanol and rinsed with de-ionized (18.2 M$\Omega$) water. Samples were mounted onto cleaned glass slides (to supply support for the thin polymer films) using double sided tape, and then coated with 10 nm of 95% Sn : 5% Sb alloy using an IBM thermal evaporator. To maintain alloy stoichiometry, fresh boats and alloy were used for each evaporation. Film thickness was measured using an Infinicon XTM-2 quartz crystal thickness sensor. The ion implanted group was then exposed to a 50 keV, 7° incident $N^+$ diffuse ion beam to a dose of $1 \times 10^{15}$ or $1 \times 10^{16}$ ions/cm$^2$ using an IBM Taconic ion implanter.

**Contacting:** Samples were pre-cleaned using a methanol rinse followed by drying with $N_2$ gas. Electrical contacts were then deposited by thermal evaporation through a shadow mask using an Edwards Auto 500 system at a chamber pressure of <$5 \times 10^{-6}$ mbar. The contacts consisted of 50 nm of Ti over-coated with 50 nm of Au, both evaporated at a rate of 0.3–1 nm/s. Four contacts were deposited, each was 5 mm in diameter and centred at one of the four corners of the 15 mm square sample, creating a quasi-Van der Pauw measurement configuration (see Fig. 1(a) in the paper for a photograph of the sample). After contact deposition, the samples were removed from the evaporator and mounted on 25 mm square glass slides with double sided tape. Polyurethane-insulated copper wires (0.2 mm dia.) were attached using low melting point InAg solder. The wires were secured to the corners of the glass slide using 5-minute Araldite, and the free ends were connected to the cryostat wiring using standard SIL pin connectors.

**Electrical Measurements:** An Oxford Instruments Variable Temperature Insert (VTI) system equipped with a 10 T superconducting solenoid was used to control the sample temperature between 1.2 K and 250 K. The samples were mounted on the sample rod with their surface perpendicular to the solenoid axis and were connected to a room temperature break-out box using 40 AWG Manganin 290 wires. The samples were measured in the VTI chamber (typical environment was ~5 mbar He gas for $T > 4.2$ K and liquid He for $T < 4.2$ K) and temperature was measured using a calibrated Lakeshore CERNOX resistor mounted on the VTI heat exchanger below the sample, and cross-checked using a second calibrated CERNOX mounted in a copper spool above the sample and in close thermal contact to the sample wires. Electrical measurements were performed in both 2 and 4 terminal configuration using a Keithley 2400 source-measure unit and a Keithley 2000 Multimeter.

**Scanning Transmission Electron Microscopy (STEM):** Structural and chemical information as a function of depth was obtained by imaging cross-sectional samples prepared from the bulk material. The cross-sections (~100 nm thick, sliced normal to the surface) were prepared using a Leica Ultracut S Cryoultramicrotome and a glass knife. Dark field and bright field transmission images, and EDX line spectra were obtained using a Technai20 FEG STEM with an electron accelerating voltage of 200 kV and a beam current of 0.092 mA. The EDX line spectra were analyzed using the Technai Imaging and Analysis software package.